# Band Gap Renormalization at Different Symmetry Points in Perovskites

Lijie Wang[1,2], Razan Nughays[2], Jun Yin[3], Chun-Hua Shih[4], Tsung-Fang Guo[4], Omar F. Mohammed[2,5*] & Majed Chergui[1,6*]

[1]*Laboratory of Ultrafast Spectroscopy, ISIC and Lausanne Centre for Ultrafast Science (LACUS), École Polytechnique Fédérale de Lausanne (EPFL), CH-1015 Lausanne, Switzerland.*
[2]*Advanced Membranes and Porous Materials Center (AMPM), Division of Physical Science and Engineering, King Abdullah University of Science and Technology (KAUST), Thuwal 23955-6900, Kingdom of Saudi Arabia.*
[3]*Department of Applied Physics, The Hong Kong Polytechnic University, Kowloon 999077, Hong Kong, P. R. China.*
[4]*Department of Photonics, National Cheng Kung University, Tainan 701, Taiwan ROC.*
[5]*KAUST Catalysis Center, Division of Physical Sciences and Engineering, King Abdullah University of Science and Technology (KAUST), Thuwal 23955-6900, Kingdom of Saudi Arabia.*
[6]*Elettra Sincrotrone Trieste, Strada Statale 14 - km 163,5, 34149 Basovizza, Trieste, Italy.*

**Abstract**: Using ultrafast broad-band transient absorption (TA) spectroscopy of photo-excited MAPbBr$_3$ thin films with probe continua in the visible and the mid-to-deep-UV ranges, we capture the ultrafast gap renormalization at the fundamental gap situated at the R symmetry point of the Brillouin Zone (BZ) and a higher energy gap at the M point. Global Lifetime analysis, Lifetime density distribution and spectral trace fitting analysis are applied to extract quantitative information. Our work confirms the similarity of the energy gap renormalization at both symmetry points, which rises within the instrument response function (IRF, ~250 fs) and decays in ~400-600 fs times, and an energy red-shift of ~90-150 meV. The ability to monitor different high symmetry points in photoexcited perovskites opens exciting prospects into the characterization of materials, which can be extended to a larger class of materials.

Semiconductors are known to exhibit large changes of their optical properties in the region of the fundamental band-gap upon photoexcitation [1,2]. These changes are due to the creation of electron-hole pairs, which modify by band filling, the intra-band and inter-band contributions to the complex optical dielectric function. This is manifested in various concurrent processes, such as inter- and intra-band absorption, Pauli blocking, and band-gap renormalization (BGR) [3–5]. The latter refers to the shrinkage of the fundamental electronic band gap, and is one of the first events after photoexcitation. Understanding it holds great importance for applications in photonics, ultrafast optical switching, and even unconventional superconductivity [6–8]. BGR is commonly reported at the fundamental absorption band-edge, which in most semiconductors occurs typically in the visible spectral region [9–12]. However, a more complete characterization of materials properties and response to photoexcitation implies probing the band-edge renormalization (BER) of higher energy band-edges, which originate either from higher (deeper) conduction/valence bands (CB/VB) or from band-gaps at different symmetry points of the Band Structure diagram. This aspect has rarely been reported. It necessitates monitoring a wider wave-vector space of the BS diagram of the material. Angle-resolved Photoelectron spectroscopy (ARPES) at extreme ultraviolet energies is ideal in this respect[13], but in the absence of chemical- [14,15] or photo-doping [16,17], it monitors only the VB structure. In addition, even in the case of chemical- or photo-doping, only the valley of the CB that contain electrons can be visualized. It is however seldom the case that more than the lowest CB minimum is populated. Using time-resolved ARPES [18,19], in their study of photo-excited black phosphorus, Roth et al [16] reported the temporal evolution of the CB and VB at the Γ point of the BS diagram of the material. However, for estimating the BGR energy shift, they only focused on the VB maximum.

The alternative approach is to perform transient absorption or reflectivity (TA/TR) studies using (a) broadband continuum probe(s) that would embrace an as wide as possible observation window in energy space. Of course, this approach does not provide an observable in k-space as ARPES does, but with the help of the BS diagram, one can make the contact between the k- and energy-spaces. This approach was implemented in the case of the indirect band-gap (BG) material, anatase $TiO_2$, using a broadband mid-to-deep ultraviolet (UV) continuum probe [15], reporting transitions in different regions of the BS diagram, though the gap renormalization was not specifically investigated. This approach provide the additional advantage of cancelling out scattering effects and other contributions, by taking the difference between pumped and unpumped absorption spectrum [20], which is of importance, as the absorption spectra of semiconductors beyond the optical gap are often quasi-continuous. Here, we implemented the same approach in the case of one of the most promising optoelectronic materials, the $MAPbBr_3$ perovskite. The choice of $MAPbBr_3$ is motivated by the fact that it has been richly investigated, especially at the fundamental BG in the visible spectral range (Figure S1), and BGR is also well-documented [12,21–26]. In addition, its absorption spectrum (Figure S1a) shows modulations above the fundamental band gap (VB1 → CB1, ~2.3 eV labelled 1), which correspond to at least three edges representing transitions from the upper VB at the M and X high symmetry points to the lowest CB (labelled 3 and 4: VB1 → CB1, ~3.8 eV and VB1 → CB1, ~4.5 eV), or from a VB sub-band to the lowest CB at the R point (labelled 2: VB3→CB1 at ~3.4 eV), based on the BS diagram (Figure S1b)[34,35]. By comparing the BGR with the gap renormalization at different symmetry points (hereafter called Band-edge renormalization, BER), we can quantify the way photo-excited charge carriers affect the system in different regions of the Brillouin Zone (BZ). In the present case, we deal with the fundamental gap at the R point and the band edge at the M point of the BZ, under identical excitation conditions.

To analyse the early-time photo-induced data, we employed a global lifetime analysis (GLA) method (see SI for the method), which yields decay-associated spectra (DAS) [29,30]. Notably, DAS1, representing the spectral amplitude for the sub-ps lifetime component, exhibits a distinct derivative-like response characterized by positive signals on the low-energy side and negative signals on the high-energy side at ~3.8 eV, resembling the spectral feature due to BGR observed in the visible region. Furthermore, we determined the lifetimes and energy red-shift of BER in MAPbBR$_3$ material to be in the range of ~400-600 fs, and 90±30 and 150±40 meV at the R and M points, respectively. These results confirm that the photo-induced transient gap shrinkage occurs in both low- and high-energy bands at different symmetry points, and exhibit nearly identical behaviours at the edge of the lowest CB. This indicates that the energy shift at different symmetry points upon photo-excitation in perovskites should be considered in their device design and applications. The experimental procedures and methods are described in the Supplementary Information (SI).

**Fig. 1** shows a schematic diagramme depicting the detection of different band transitions in the MAPbBr$_3$ perovskite material using both visible and deep-UV probes. Upon absorption of photons at the pump photon energy (3.10 eV), a non-equilibrium carrier population is generated, leading to a shrinkage in the energy gap in fs timescale. The BGR at the R point (corresponding to the fundamental BG, see **Fig. S1**) is detected using a visible probe, while the high-energy inter-band transition at the M point (labelled 3 in **Fig. S1a**, at ~3.8 eV, and in **Fig. S1b**) can only be detected using a broadband UV probe. Since our continuum probe covers the 3.3 to 4.3 eV range, this makes the detection of BER at transitions 2 and 4 unfeasible, and we therefore limit our study to transition 3 for the higher energy edge.

The MAPbBr$_3$ thin films were first investigated in the visible spectral region, encompassing the transition of VB1 → CB1 at the R point (fundamental BG transition), as extensively studied previously[12,21,23]. **Fig. 2a** shows the time-energy TA map in a 0-5 ps time window of the MAPbBr$_3$ perovskite excited at 3.10 eV, and **Fig. 2b** shows the corresponding spectral traces at various delay times, revealing distinct negative and positive features from low to high energies, along with a small positive shoulder on the lower-energy side at 0.2 and 0.4 ps. These figures display the characteristic bleaching at the optical BG transition at ~2.35 eV (see **Fig. S1**), accompanied by a broad, weak absorption signal on the higher-energy side. The transient features observed in the visible region have been extensively discussed in previous studies [12,26,31–36]. However, a small positive signal promptly emerges on the lower-energy side of the BG (indicated by the dashed red circle in **Fig. 2a**), which we attribute to the reduction of the BG caused by the presence of hot carriers on timescales of tens to hundreds of fs [12], namely BGR. As the excited hot carriers occupy fewer states at the new, lowered band edge compared to colder-carrier distributions, the TA signal at the renormalized band edge displays a photo-induced absorption (PIA), which disappears on a sub-ps timescale as the carrier temperature rapidly decreases [12].

We conducted a GLA analysis that simultaneously examines multiple kinetic traces recorded at different probe energies (see SI for the method). **Fig. 2c** presents the TA map retrieved from the extracted DAS, which successfully captures all the spectral features observed in **Fig. 2a**. The resulting DAS associated with two different lifetimes, are depicted in **Fig. 2d**, and correspond to an initial sub-ps process (DAS1) and a long-term (> 5 ps) evolution process (DAS2). DAS1, exhibits a response resembling a derivative-like shape, with positive values on the low-energy side and negative ones on the high-energy side, as expected for the BGR and its immediate red-shift after photo-excitation. On the other hand, DAS2 shows the shape of the photo-induced electronic signal at longer timescale (**Fig. 2b**).

We now turn to the band-edge response at the M-point. **Fig. 3a** presents the time-energy TA map in the 3.3 to 4.3 eV region, under identical pump conditions as for the experiments in the visible range. Additionally, **Fig. 3b** displays the corresponding spectral traces obtained from the UV-probed measurements. The transients exhibit three pronounced negative bands at approximately 3.4 eV, 3.8 eV, and above 4.3 eV, accompanied by a positive signal around 4.15 eV. The assignment of these features is supported by steady-state absorption measurements of a MAPbBr$_3$ single crystal and ellipsometry data shown in **Fig. S1a and b** (labelled transition 2: VB3→CB1 at ~3.4 eV, and transition 3 and 4: VB1 → CB1, ~3.8 eV and VB1 → CB1, ~4.5 eV, see **Note S1**, SI).

Notably, there is a weak positive signal at around 3.6 eV, which is close to zero and disappears within approximately 0.5 ps (indicated by the dashed red circle in **Fig. 3a**). This behaviour is also visible in **Fig. 3b**. Unlike the immediate negative response observed at 3.4 and 3.8 eV after photoexcitation (e.g., at 100 fs), the signal at 3.6 eV remains close to zero before the negative component increases sharply. This delayed increase of negative amplitude is particularly evident in the early temporal traces shown in **Fig. 3c**. In contrast to the TA signal at 3.4 and 3.5 eV, the 3.6 eV signal exhibits a delayed signal increase of ~0.3 ps. Additionally, this short-lived signal appears slightly below the 3.8 eV band, which corresponds to the energy gap transition at the M symmetry point in the BZ. However, the strong superposition of negative signals may obscure the nature of the positive response at around 3.6 eV. Nevertheless, to unravel the underlying spectral components at different delay stages, a GLA was performed to generate the DAS that provide the spectral components at each lifetime. **Fig. 3d** presents the reconstructed TA map obtained from the GLA, while **Fig. 3e** shows the resulting DAS1 and DAS2. The DAS2 basically resembles the long-term spectral responses. Surprisingly, DAS1, representing the spectral amplitude of the sub-ps lifetime component, exhibits a similar BGR derivative-like profile as in the visible region, suggesting a BER at the M symmetry point (see **Fig. S1b** and **Note S2**, SI). Since the excitation photon energy is lower than the energy gap at the M point (~3.8 eV), the observed bleaching (negative) signal and BER effect at this symmetry point are likely attributed to the redistribution of photo-induced electron and hole populations (Coulomb interactions).

To determine the lifetime of the BER at ~3.8 eV, we performed a lifetime density distribution (LDD) analysis that compresses the kinetic information into a distribution map (see SI for the method). **Fig. 4a** presents the LDD map, which extends up to a lifetime of 100 ps and is fitted to the ΔA data depicted in

**Fig. 3a**. Gaussian shapes were used to simulate a three-dimensional lifetime distribution map that encompasses both spectral and lifetime distributions. Irrespective of the probe energy, the LDD map exhibits distantly spaced distributions, with a notable sub-ps peak discernible in the time domain spanning the energy range of 3.4-4.2 eV. Since the LDD map serves as a quasi-continuous analogue of the DAS, the amplitudes observed at shorter lifetimes (e.g. <1 ps) exhibit a similar profile to DAS1. Meanwhile, the amplitudes at longer lifetimes (e.g., 10-100 ps) resemble DAS2, alongside the spectral traces observed during the long-term decay. This agreement further validates the reliability and reproducibility of the GLA and LDD fitting. **Fig. 4b** showcases a lifetime trace taken at 3.8 eV, with the first peak (L1 in the figure) at ~530 fs, which represents the most probable first time constant value. Taking into account the variations in lifetime values observed for different detection energies around 3.8 eV (as shown in **Fig. S2,** ~420 fs at 3.6 eV, ~620 fs at 3.7 eV, and ~600 fs at 3.9 eV), it can be concluded that the BER effect at this high-energy band exhibits a lifetime range of 400-600 fs, which aligns well with the BGR in the visible spectral region (see **Fig. S3,** and summarized parameters in **Table S1**).

Since the immediate photo-induced TA responses at the M symmetry point arise from a combination of inter- and intra-band electronic signals and BER effects, it is essential to quantitatively assess their relative contributions and temporal evolution. To achieve this, we first normalize the transient spectra at 3.8 eV in the early time regime (**Fig. S4**). The spectral evolution observed at ~3.6 eV and ~4.1 eV clearly exhibits distinct behaviours before and after 300 fs, indicating the presence of two different responses with different lifetimes. We reconstructed the differential absorption map using a ratio number $n$, to simulate the complex spectral evolution following photoexcitation. The signal amplitude was calculated based on DAS1 + $n\times$DAS2, with $n$ varying from 1 to 8 to represent the decreasing spectral weight of the BER component. Remarkably, the reconstructed differential absorption map shown in **Fig. 4c** faithfully reproduces the experimental TA profiles and captures their temporal evolution (see **Fig. S5** for the comparison). We selected $n = 1.7$ and 5 (see **Fig. S6**) to represent the early and later delay times, and the reconstructed absorption changes exhibited great similarities with the TA spectral traces recorded at 100 fs and 600 fs. Notably, the variations in negative peaks around ~3.4 and ~3.8 eV were also accurately reproduced. Given the role of DAS1 as an indicator of BER, this provides an opportunity to fit the early time spectral traces (<1 ps) using various combinations of DAS1 and DAS2. **Fig. 4d** displays the experimental UV-probed spectral trace at 0.1 ps, along with its fitting using combinations of a×DAS1 + b×DAS2. The fitted curve well replicates the experimental data that is represented by the grey solid line, and the relative contribution of BER to the overall TA signal (BER/TA) is determined by a/(a+b)×100%. Moreover, **Fig. S7** presents spectral trace fittings within the first 1 ps for 0.1 ps increments, and the relative contributions of BER effect constantly decrease from 42% at 0.1 ps to ~6.5% at 1 ps, indicating a negligible spectral component of BER effect after a delay time of 1 ps.

**Fig. 4e** presents a schematic diagram illustrating the generation of the derivative-like spectral shape through a Gaussian-like spectral peak redshift. In a reverse manner, we fit the DAS1 obtained from the UV probing using two subtracted Gaussian functions and determine a peak redshift of 150 ± 40 meV (see **Fig. 4f**). This value aligns closely with the observed fundamental BGR in the visible spectral region (see **Fig. S8** and **Tables S2**) and falls in the same level as the reported photo-induced BRG timescale [37], and can be considered to demonstrate the displacement of the band edge.

In summary, we conducted broadband TA spectroscopy spanning from the visible to mid-to-deep-UV ranges on MAPbBr$_3$ thin films, with GLA, LDD, and spectral trace fittings analysis. The use of probe continua enables the investigation of excitations at various symmetry points within the BZ, offering a way of accessing them, which would otherwise not be possible using other methods (e.g. ARPES). Our results provide invaluable insights into the depth of kinetic information embedded within the experimental data, and demonstrate that the transient gap shrinkage is identical at the R and M symmetry points. This is not unexpected since in both cases, we are monitoring the edge of the lowest conduction band. A further extension of this work could involve monitoring the response of (deeper) higher sub-bands of the (valence) conduction band, e.g. in the present case at transitions 2 and 4. Uncovering such higher energy BER's, which has not been previously considered, would shed light on the fundamental transient physical behaviours under non-equilibrium conditions in perovskite materials. These findings would also provide support for investigating underlying mechanisms in other materials, such as ultrafast optical switching and unconventional superconductivity, behind carrier-induced bands renormalization [8,38,39].

This work was supported by the Swiss NSF via the NCCR: MUST, the European Research Council Advanced Grant DYNAMOX and by the King Abdullah University of Science and Technology (KAUST).

*_omar.abdelsaboor@kaust.edu.sa_.

*_majed.chergui@epfl.ch_;

*_majed.chergui@elettra.eu_;

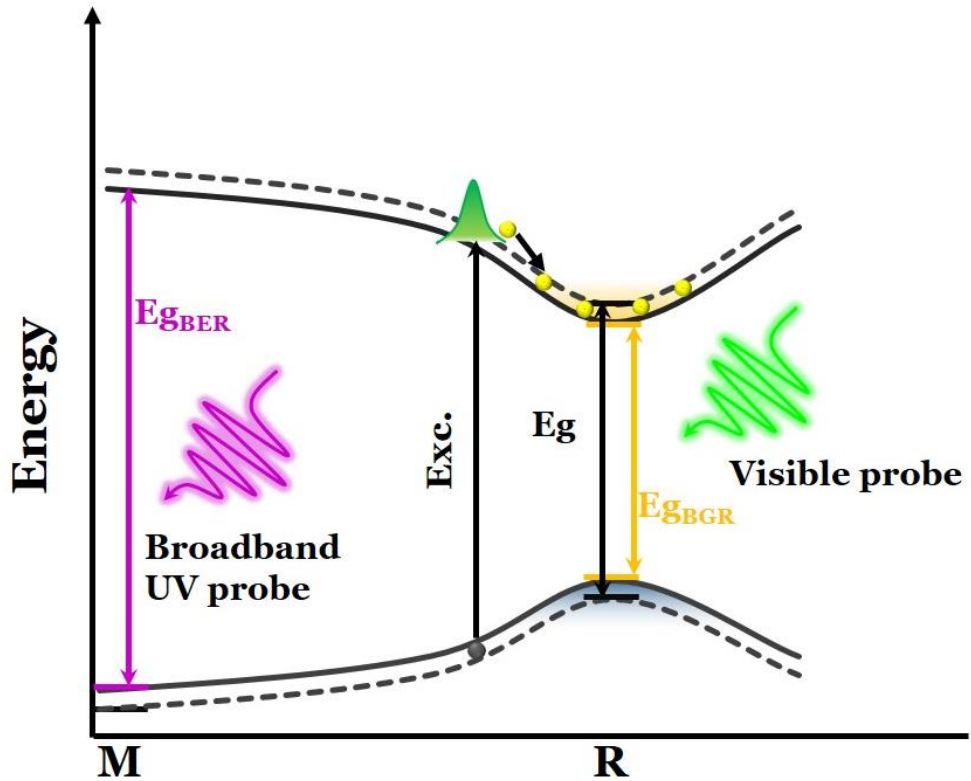

**Fig. 1** Schematic diagram of the visible and deep-UV based detection of band transitions in the MAPbBr$_3$ perovskite material. Upon photon absorption at the pump photon energy (3.10 eV), a non-equilibrium carrier population is generated, resulting in a shrinkage of the initial energy gap. The BGR occurring at the R point (corresponding to the BG) can be detected by a visible probe, whereas the high-energy BER effect at the M point is detected using a UV probe.

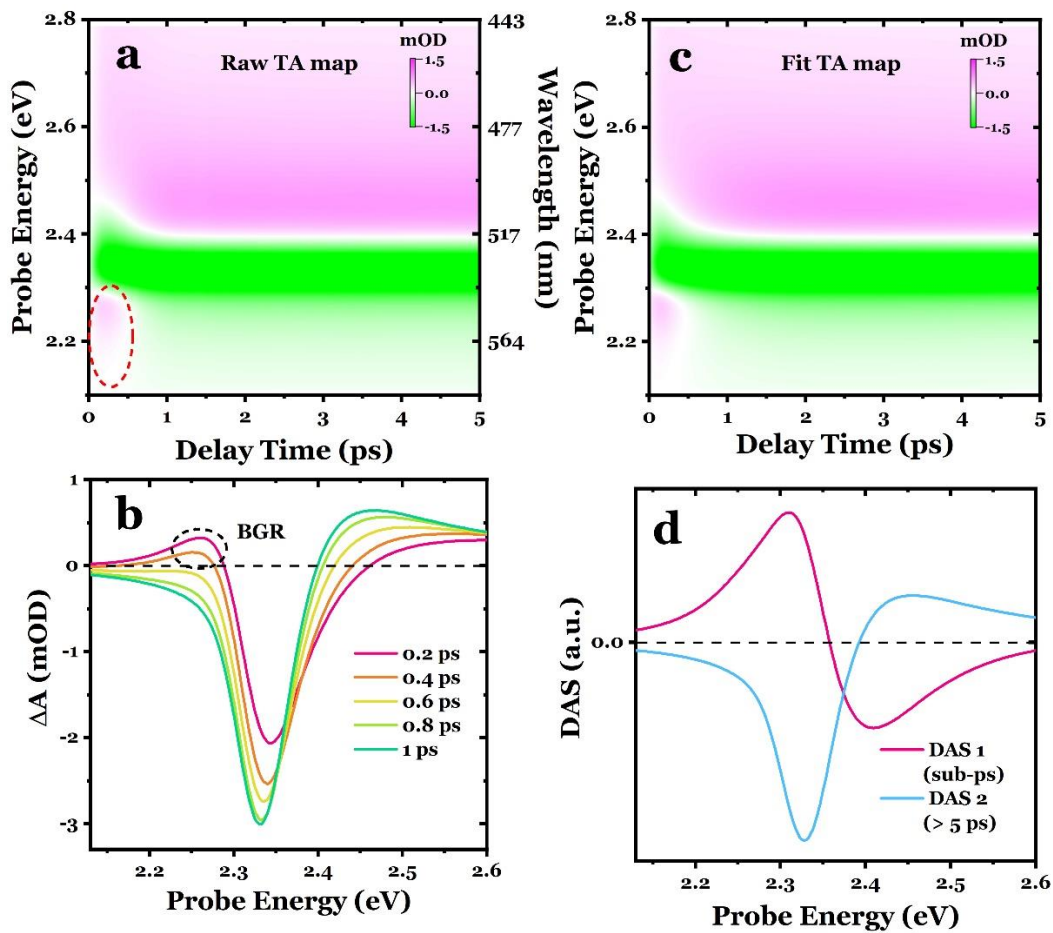

**Fig. 2 a**, Experimental TA time-energy map probed in the visible spectral region. **b**, The photo-induced transients at 0.2, 0.4, 0.6, 0.8, and 1 ps. **c**, The fitted TA map, by using global lifetime analysis method. **d**, The decay-associated spectra corresponding to a fast and a slow evolution processes. The fitting was performed within a time window of 5 ps.

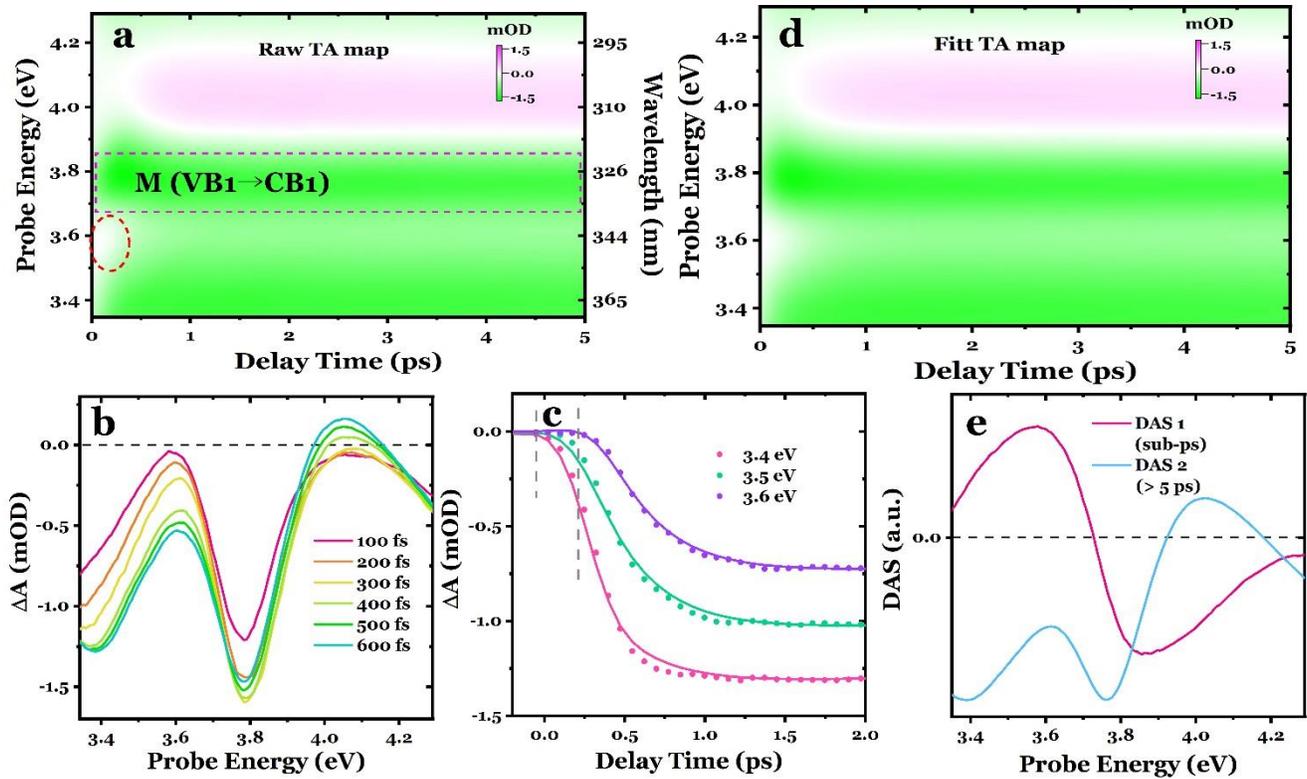

**Fig. 3 a**, Experimental TA time-energy map probed in the mid-to-deep-UV spectral region. **b**, The photo-induced transients probed in the UV spectral region, at 100, 200, 300, 400, 500, and 600 fs, respectively. **c**, The early temporal traces probed at 3.4, 3.5, and 3.6 eV, along with their fittings. **d**, The fitted TA map, by using global lifetime analysis. **e**, The decay-associated spectra that corresponding to a fast and a slow processes after photoexcitation.

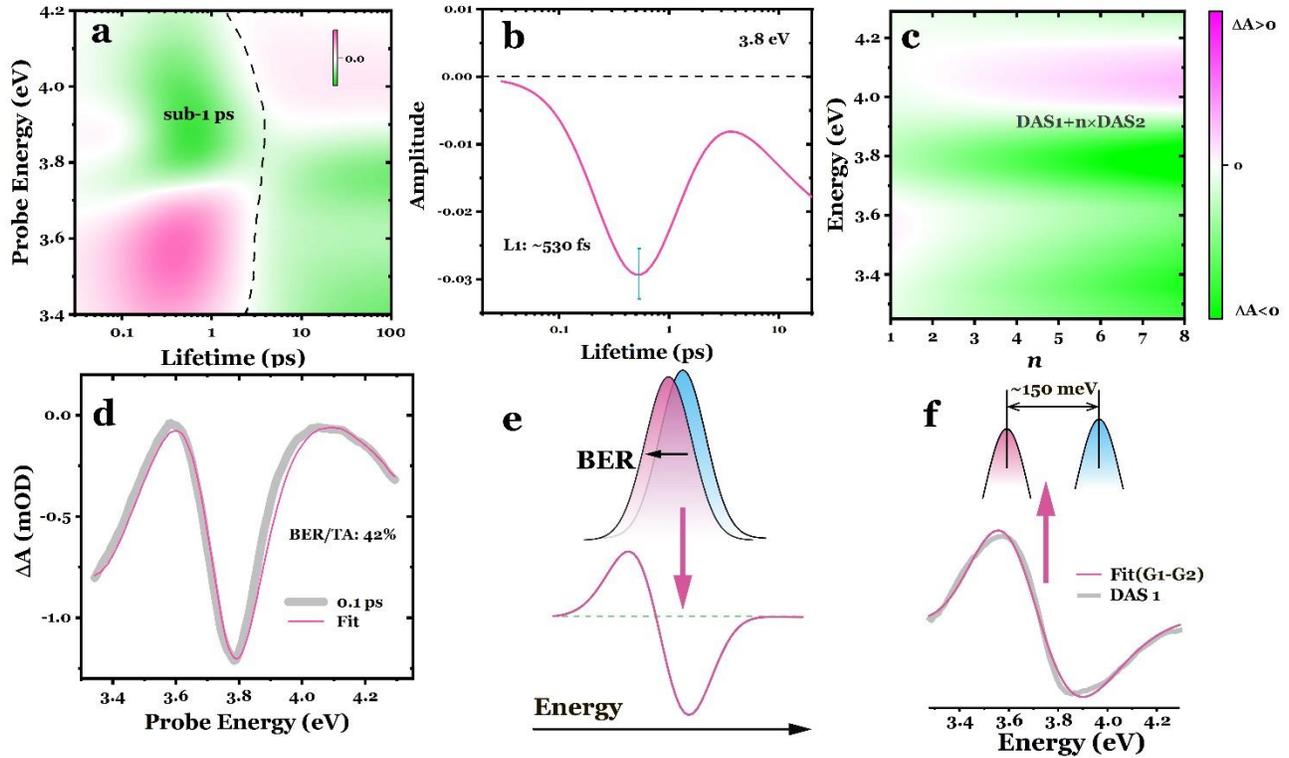

**Fig. 4 a**, Lifetime density distribution map within 100 ps, fitted to the ΔA data in Fig. 3a. The spectral amplitudes observed at shorter lifetimes (e.g. <1 ps) exhibit a similar profile to DAS1, while the amplitudes at longer lifetimes (e.g., 10-100 ps) bear resemblance to DAS2. **b**, Lifetime trace of the LDD at the probe energy of 3.8 eV, a clear peak (L1) can be resolved in the first 1 ps, with the L1 at the amplitude maximum position of ~530 fs. **c**, The re-constructed differential absorption map by using different ratio number *n*. The signal amplitude is calculated based on DAS1 + *n*×DAS2 (simulating the relative increasing contribution from DAS2), to simulate the complex spectral evolution after photoexcitation. **d**, Experimental UV-probed spectral trace at 0.1 ps and its fitting using combinations of DAS spectra (a×DAS1 + b×DAS2). The relative contribution of BER to the overall TA signal (BER/TA) is demonstrated in the figure. **e**, Schematic diagram illustrating band energy reduction (red-shift) resulting in a derivative-like transient signal, which is positive in the low-energy part and negative in the high-energy part. **f**, Fitting of the DAS1 signal using two Gaussian functions (Gaussian1 minus Gaussian2). A band red-shift of approximately 150 meV was obtained.

# Band Gap Renormalization at Different Symmetry Points in Perovskites

Lijie Wang[1,2], Razan Nughays[2], Jun Yin[3], Chun-Hua Shih[4], Tsung-Fang Guo[4], Omar F. Mohammed[2,5]* & Majed Chergui[1,6]*

[1]*Laboratory of Ultrafast Spectroscopy, ISIC and Lausanne Centre for Ultrafast Science (LACUS), École Polytechnique Fédérale de Lausanne (EPFL), CH-1015 Lausanne, Switzerland.*
[2]*Advanced Membranes and Porous Materials Center (AMPM), Division of Physical Science and Engineering, King Abdullah University of Science and Technology (KAUST), Thuwal 23955-6900, Kingdom of Saudi Arabia.*
[3]*Department of Applied Physics, The Hong Kong Polytechnic University, Kowloon 999077, Hong Kong, P. R. China.*
[4]*Department of Photonics, National Cheng Kung University, Tainan 701, Taiwan ROC.*
[6]*KAUST Catalysis Center, Division of Physical Sciences and Engineering, King Abdullah University of Science and Technology (KAUST), Thuwal 23955-6900, Kingdom of Saudi Arabia.*
[5]*Elettra Sincrotrone Trieste, Strada Statale 14 - km 163,5, 34149 Basovizza, Trieste, Italy.*

# Experimental Methods

**Sample preparation and characterization.** For MAPbBr$_3$ films, prior to the deposition, the quartz substrates underwent a sequential cleaning process. Firstly, the substrates were subjected to ultrasonic treatment in detergent, de-ionized water, acetone, and isopropyl alcohol. After drying, the cleaned substrates were further treated with UV-ozone (Model: 42, Jelight, USA) for 25 minutes. The MAPbBr$_3$ precursor solution was prepared by combining 0.015 g of MABr (Dyesol) with 0.0481 g of PbBr$_2$ (Sigma-Aldrich, 99.999%) in an anhydrous dimethyl sulfoxide (DMSO, Sigma-Aldrich) solution (1070 μl) at 60°C. The solution was stirred for 12 hours. To fabricate the MAPbBr$_3$ perovskite thin film, a consecutive two-step spin-coating process was employed. The solution was spin-coated onto the quartz substrates at 500 rpm for 7 seconds, followed by a spin-coating at 4000 rpm for 70 seconds. Additionally, at 43 seconds during the spin coating, 250 μl of chloroform solvent was dropped onto the surface of the precursor film. Subsequently, the MAPbBr$_3$ perovskite film was annealed on a hotplate at 70°C for 10 minutes. It is noteworthy that all the procedures for preparing the MAPbBr$_3$ precursor solution and films were conducted inside a nitrogen-filled glove box with oxygen and moisture levels maintained below 1 ppm.

For MAPbBr$_3$ single crystals used for ellipsometric experiments, the precursor MABr (0.748 g) was dissolved in anhydrous dimethylformamide (4 mL) in a 20 mL glass vial to form a clear solution. Then, PbBr$_2$ (2.452 g) was added into the glass vial with stirring to obtain a nearly saturated clear MAPbBr$_3$ solution. The glass vial was then placed onto a hotplate at 50 °C without disturbance for slow evaporation. Bulk MAPbBr$_3$ single crystals with dimensions in the centimeter range can be obtained from the solution after 12 h. These procedures were all performed inside a fume hood.

**Transient spectroscopic measurements.** The experiments were performed using two different setups: one for broadband visible probe and the other for broadband deep-UV probe.

(a) For the broadband visible probe setup, a 1 kHz regenerative amplifier provides 30 fs pulses at 800 nm with an energy of approximately 720 µJ per pulse. A noncollinear optical parametric amplifier (NOPA) was utilized to generate tunable visible pump pulses with ~15 nm bandwidth and energy ranging from 2–4 µJ per pulse. The probe beam was focused onto a $CaF_2$ plane to generate white light in the range of 450–750 nm (1.65–2.75 eV).

(b) For the broadband deep-UV probe setup [1,2], a 20 kHz Ti:sapphire regenerative amplifier (KMLabs, Wyvern500), providing 50 fs pulses at 800 nm with an energy of 0.6 mJ. These pulses were used to pumps a NOPA, generating sub-90 fs visible pulses at 13 µJ per pulse in the range of 510–740 nm (1.68–2.43 eV). About 40% of the NOPA output was used to generate broadband UV probe pulses with a bandwidth of ~100 nm through an achromatic doubling scheme [3]. The probe pulses were further compressed using chirp mirrors and was determined with a commercial FROG system (Swamp Optics) to be <20 fs pulse duration. The relative polarization between the pump and probe beams was set at the magic angle (54.74°) using a half-wave plate to avoid photo-selection effects. After passing through the sample, the transmitted broadband probe beam was focused into a 5 m multi-mode optical fiber, which was coupled to the entrance slit of a 0.25 m imaging spectrograph (Chromex 250is). The beam was dispersed by a 150 gr/mm holographic grating and imaged onto a multichannel detector consisting of a 512-pixel CMOS linear sensor (Hamamatsu S11105) with a pixel size of 12.5×250 µm. The pixel readout rate could reach up to 50 MHz. The typical spot sizes of the pump and probe beams were approximately 120 µm and 50 µm full widths at half-maximum, respectively.

In all measurements, the pump fluence at 400 nm (3.1 eV) was approximately 50 µJ/cm$^2$ with ~10% uncertainty due to laser power measurement and laser beam spot size. The pump power was recorded on a shot-to-shot basis using a calibrated photodiode for each pump wavelength, enabling the normalization of the data for the pump power. The instrument response function (IRF) was determined by measuring the cross phase modulation (CPM) signal at time zero of the pure quartz substrate, and was found to be ~250 fs. The thin film perovskite samples were mounted in a film sample holder with a nitrogen gas flow to protect the sample surface. The probe signal was measured after transmission through the sample, and its detection was synchronized with the laser repetition rate.

**Spectroscopic ellipsometry.** Spectroscopic ellipsometry was performed using an M-2000 DI device (J. A. Woollam, USA), which operated in the 193–1690 nm wavelength range. The sample was measured at a minimum of three angles of incidences (65°, 70°, and 75°), and the data analysis was performed using the Complete EASE 6.51 software package, to generate the absorption coefficient of perovskite signal crystal.

**Band structure calculations.** We performed the BS calculations using the projector-augmented wave method implemented in the Vienna Ab Initio Simulation Package code [4,5]. The GGA and PBE exchange-correlation functional were used, and van der Waals interactions were also included in the calculations using the zero-damping DFT-D3 method of Grimme. A uniform grid of 6 × 6 × 6 k-mesh in the Brillouin zone was employed to optimize the crystal structure of cubic-phase MAPbBr$_3$. The energy cutoffs of the wave functions were set to 500 eV for bulk MAPbBr$_3$.

**Global analysis.** We conducted a GLA analysis that simultaneously examines multiple kinetic traces recorded at different probe energies, using a discrete sum-of-exponentials function [6]:

$$S(t, \lambda_{exc}, \lambda_{pro}) = \sum_{j=1}^{n} A_j(\tau, \lambda_{exc}, \lambda_{pro}) \exp\left(-{t}/{\tau_j}\right) \otimes IRF(t) \tag{1}$$

Where, $\tau$, represent the global lifetimes, and A, are the amplitudes for each kinetic trace. The detected signals are convoluted with the IRF, which is modeled by a polynomial function [7]:

$$IRF(\lambda) = c_0 + \sum_{i=1}^{n} c_i \left(\frac{\lambda - \lambda_c}{100}\right)^i \qquad (2)$$

The time zero position at the central wavelength, $\lambda_c$, is given by $c_0$. GLA yields the so-called DAS, where the pre-exponential amplitudes for each lifetime component are plotted against probe wavelength, $\lambda_{pro}$, and the DAS is a compact representation of the kinetic information in the data. Fig. 2c presents the TA map retrieved from the extracted DAS, which successfully captures all the spectral features observed in Fig. 2a.

**Lifetime density distribution analysis.** To determine the lifetime of the BER effect at ~3.8 eV, we performed a lifetime density distribution (LDD) analysis that compresses the kinetic information into a distribution map (see SI for the method). The great number of spatial, energetic, and temporal degrees of freedom of the ultrafast responses and their matrix produces a continuous distribution of individual exponential decays [6]:

$$S(t, \lambda_{exc}, \lambda_{pro}) = \int_0^\infty \Phi(\tau, \lambda_{exc}, \lambda_{pro}) \exp(-t/\tau) d\tau \qquad (3)$$

The function $S(t)$ represents the Laplace transform of the spectral distribution function, $\Phi(\tau)$ [8]. The integral in eq (3) needs to be discretized into a quasi-continuous sum of $n$ exponential functions similar to eq (1) but with $n$ typically >50. Thus the LDD analysis offers a comprehensive overview of the kinetics, allowing for the resolution of complex analysis issues, such as non-exponential or stretched exponential kinetics [6].

**Supplementary Note S1: The band assignments in the visible-to-deep-UV probe region**

Fig. S1a displays a typical absorption spectrum of MAPbBr$_3$ material, revealing a distinct excitonic feature appears at ~2.3 eV, accompanied by multiple absorption peaks at approximately 3.4 eV, 3.8 eV, and 4.45 eV. These features can be attributed to specific transitions based on the calculated band structure diagram presented in Fig. S1b and they are annotated in the absorption spectrum. Specifically, they are identified as VB1 → CB1 at the R point (transition 1), corresponding to the BG transition in the visible spectral region; and VB3 → CB1 at the R point (transition 2), VB1 → CB1 at the M point (transition 3), and VB1 → CB1 at the X point (transition 4), respectively [9,10] in the mid-to deep-UV region.

The transient spectral traces of MAPbBr$_3$, probed in the visible and UV ranges, are depicted in Fig. 2a and 3a. The spectra exhibit distinct negative peaks located at approximately 2.35, 3.4 eV, 3.8 eV, and >4.3 eV. The energies of these bleaching (negative) signals align well with the ellipsometric measurements conducted within the same spectral range. Following a similar methodology used for the assignment of interband transitions in MAPbI$_3$ [9,11], we determined the energy distances at each symmetry point in the Brillouin zone (BZ). Consequently, the ~2.35 eV peak is assigned to the transition from VB1 to CB1 at the R point, the ~3.4 eV peak is assigned to the transition from VB3 to CB1 at the R point, while the ~3.8 eV and ~4.45 eV peaks are assigned to transitions between VB1 and CB1 at the M and X points, respectively (Fig. S1b).

When excited below the energy gap at the M and X points, direct population at these high-symmetry points is not involved. Therefore, the transient signal at ~3.4 eV (R point) exhibits different sensitivity compared to the bleaches at ~3.8 eV (M point) and ~4.45 eV (X point). This distinction is apparent in the early time traces shown in Fig. S9, where the transient signals at ~3.8 eV experiences a prompt rise, while the signal at ~3.4 eV rises more gradually due to the cooling of electrons towards the bottom of CB at the M point. It is also worth to note that since our continuum probe covers the 3.3 to 4.3 eV range, which makes the detection of BER at transitions 2 and 4 unfeasible, and we therefore focus our study on transition 3 for the higher energy edge.

**Supplementary Note S2: Decay associated spectra in the presence of time dependent spectral shifts**

The analysis of time resolved data is of crucial importance for resolving the underlying physical quantities. A powerful and widely adopted tool for this purpose is global lifetime analysis (GLA) [12]. In GLA, a sum of exponentials with energy/wavelength dependent amplitudes is fitted, typically by least square routines, to the measured data. The outcome of such a procedure includes not only the exponential decay times but also decay associated spectra (DAS), which represent the energy/wavelength dependent amplitudes of the respective exponential contributions.

Inherent to the GLA-approach is the assumption that the time and energy/wavelength dependence of the data are separable. However, this becomes challenging in the presence of a time dependent spectral shift, as the basic assumption of separability of time and energy/wavelength dependence does not hold [13,14].

In the GLA-fitted TA map of perovskite materials, as shown in Fig. 2 and 3, the spectral red-shift at the shortest timescales can yield an orthogonal component resembling a derivative-like shape. The DAS with a larger time constant reflects the transient signal due to photo-induced bleaching or absorption, while the DAS associated with shorter time constants, namely < 1 ps, parametrize the shift dynamics at the corresponding bands. Comparing the time constant of the DAS1 with the spectral shift time constants will elucidate whether this first process can be directly attributed to a physical process or is merely a distortion by the GLA [15]. In Fig. S10, The overall shift at around ~3.6 eV can be related to an exponential shift with a time constant of $300 \pm 100$ fs, which is in good agreement with the time constant of DAS1 (~360 fs). Additionally, the assumption of a preserved line-shape is fulfilled to a sufficient extent for the analysis presented here, and the evidence is that the full width at half maximum, for instance at around 3.6 eV, varies about 15% during the first 600 fs.

Nevertheless, it is essential to emphasize that the purpose of conducting a GLA of the time-energy maps of perovskite with a specifically 0-5 ps time window is to resolve information hidden due to spectral overlap, rather than to obtain kinetic parameters. The presence of a significant orthogonal component resembling a derivative-like shape (DAS1 in Fig. 2d and 3e) serves as evidence that even in the high-energy mid-to-deep UV probe region, there is a spectral redshift similar to that of the BG transition immediately after photoexcitation near the M point in the BZ.

**Table S1**. Comparison of time constants at different energies resulting from LDD according to Eq. (3) at the fundamental BG at the R point and the higher BG at the M point.

| At the fundamental gap (R point) | 2.3 eV | 2.4 eV | 2.5 eV | - |
|---|---|---|---|---|
| | ~530 fs | ~680 fs | ~510 fs | - |
| At the M point gap | 3.6 eV | 3.7 eV | 3.8 eV | 3.9 eV |
| | ~420 fs | ~620 fs | ~530 fs | ~600 fs |

**Table S2.** The extent of energy shift at the fundamental BG and at the M point.

| Fundamental BG | M point gap |
|---|---|
| 90 ± 30 meV | 150 ± 40 meV |

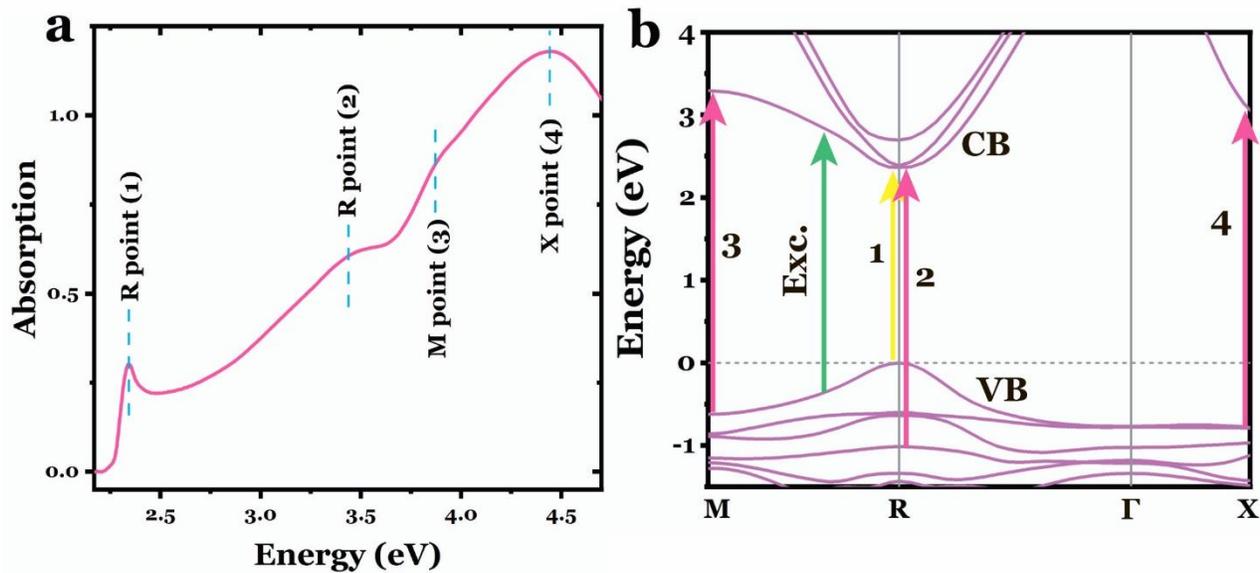

**Fig. S1. a**, The absorption spectrum of a MAPbBr$_3$ single crystal. The absorption peaks at each energies are labeled according to their corresponding interband transitions at different symmetry points. **b**, Calculated energy band diagram (the details are described in the method section). The 3.1 eV excitation is indicated by the green arrow, while the probed bleached signals that can be detected by the employed broadband visible and mid-to-deep-UV probes are represented by the yellow and red arrows, respectively. The number 1, 2 and 3 are labeled in the order of the CB and VB at each symmetry point.

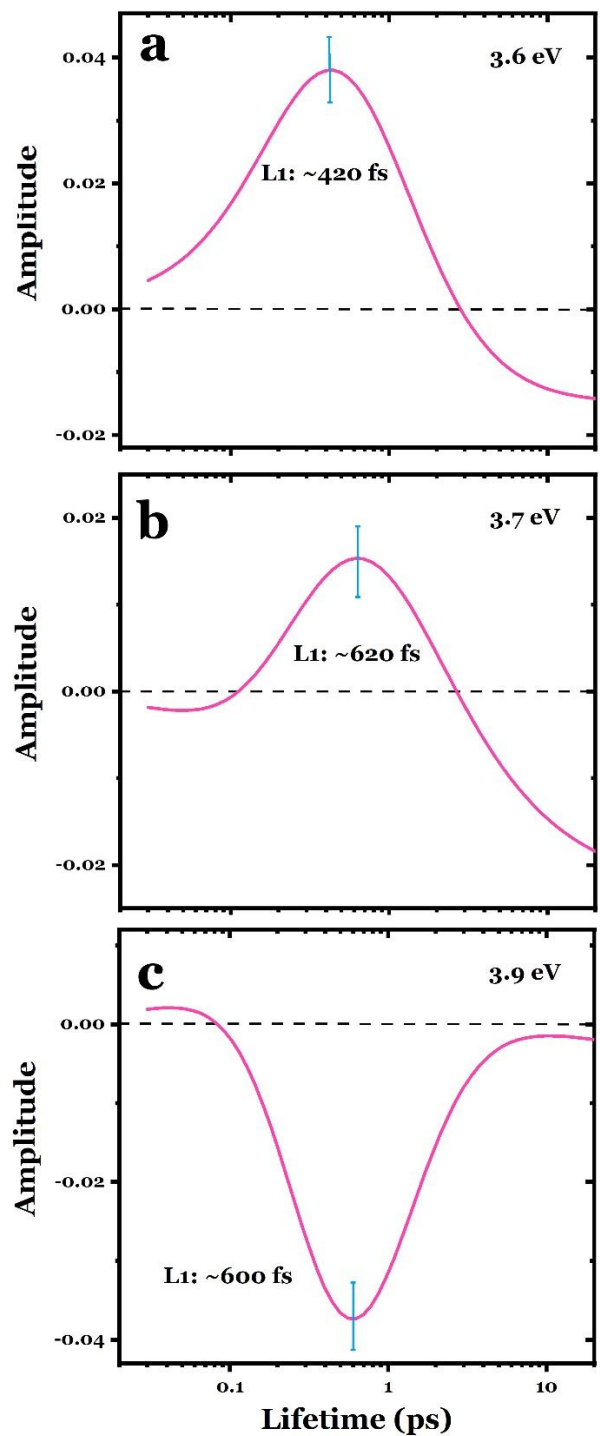

**Fig. S2.** Lifetime traces of the LDD at the probe energy of a, 3.6 eV; b, 3.7 eV; c, 3.9 eV. A distinct peak can be resolved in the first 1 ps at different probed energies. The L1 peak position of at 3.6, 3.7, and 3.9 eV are ~420, ~620, and ~600 fs, respectively.

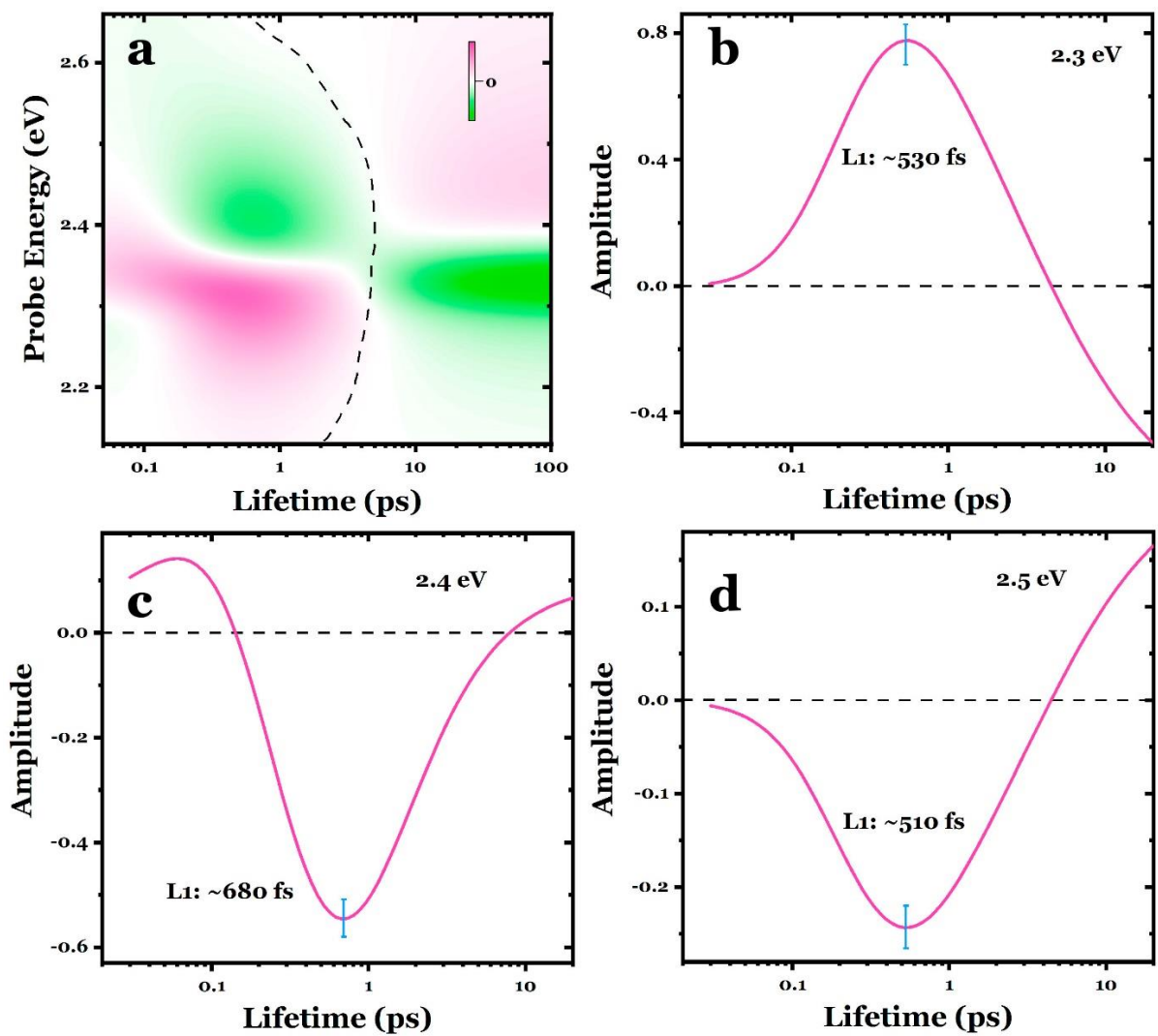

**Fig. S3. a**, Lifetime density distribution map within 100 ps, fitted to the ΔA data in Fig. 2a. **b-d**, Lifetime trace of the LDD at the probe energy of 2.3, 2.4, and 2.5 eV, a clear peak (L1) can be resolved in the first 1 ps, with the L1 at the amplitude maximum position of ~530, ~680, and ~510 fs.

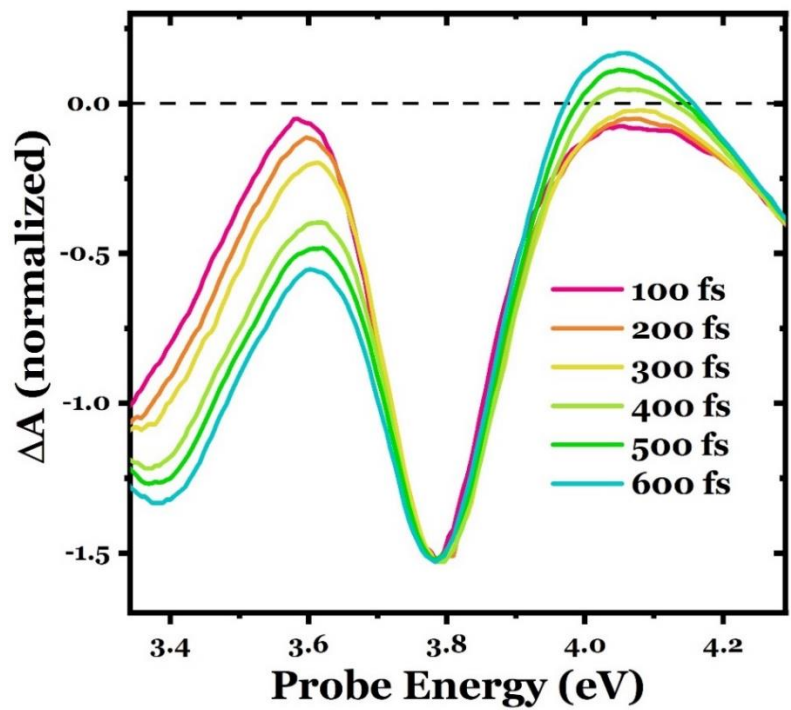

**Fig. S4.** Normalized early time transient spectra. The spectral evolution at ~3.6 eV and at ~4.1 eV clearly reveals different behaviors before and after ~300 fs. These spectra are normalized at ~3.8 eV.

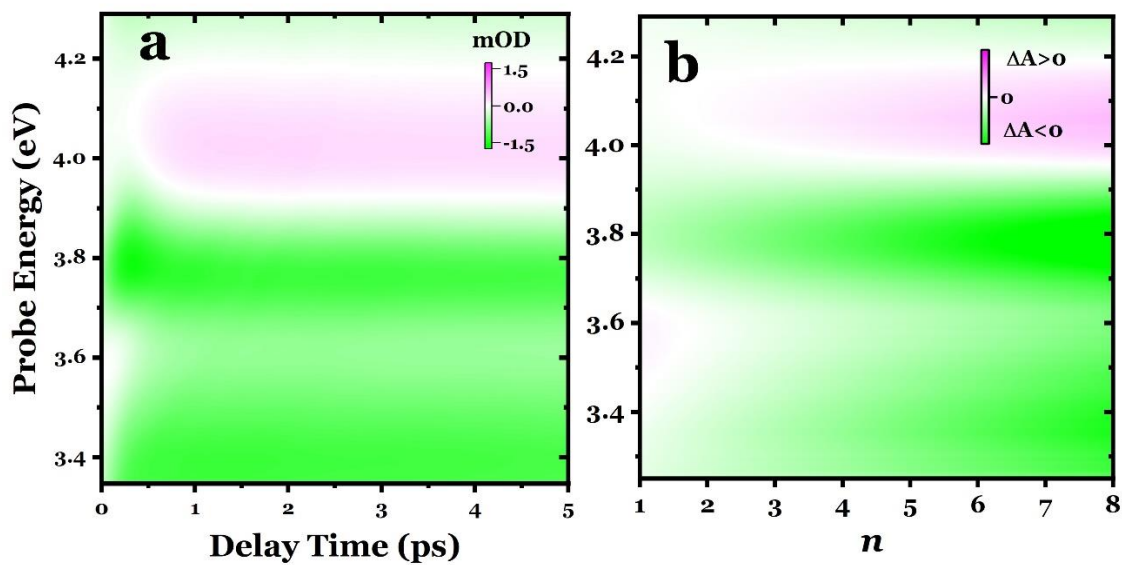

**Fig. S5. a**, Experimental TA time-energy map probed in the mid-to-deep-UV spectral region. **b,** The reconstructed differential absorption map by using different ratio number *n*. The signal amplitude is calculated based on DAS1 + $n \times$DAS2 (simulating the relative increasing contribution from DAS2), to simulate the complex spectral evolution after photoexcitation.

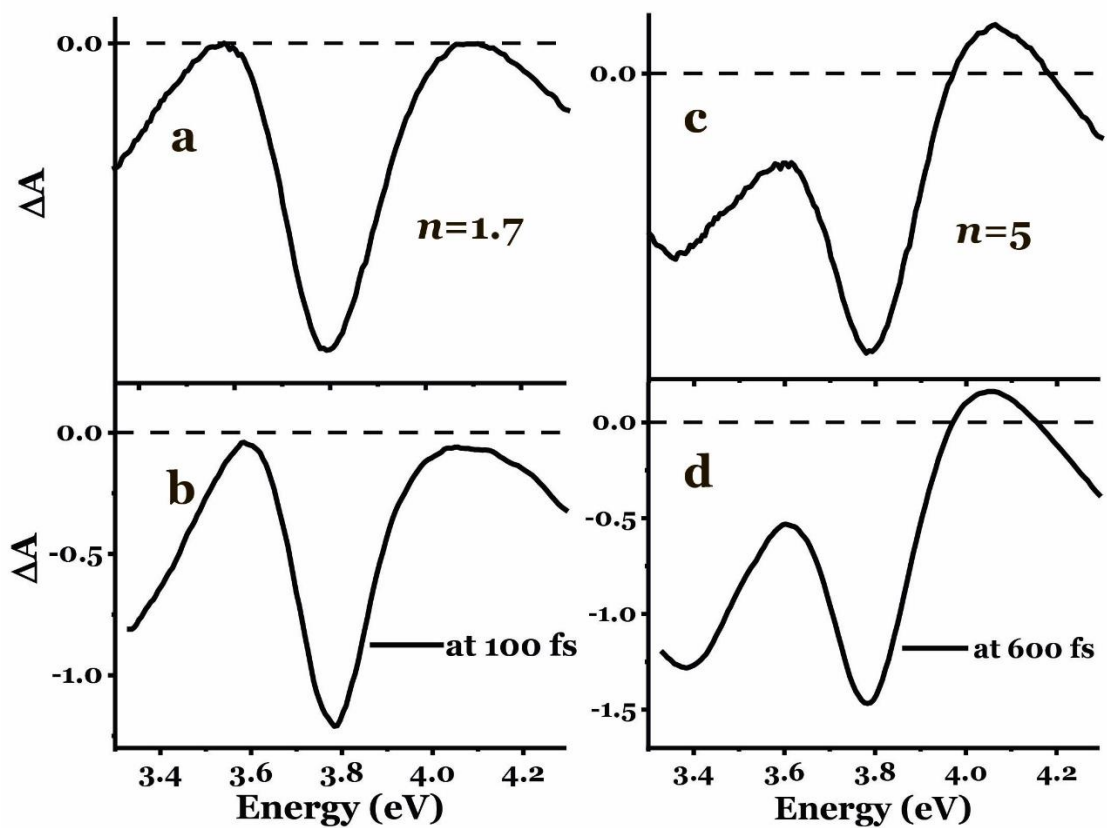

**Fig. S6.** The examples of spectral trace comparisons between *n*=1.7 and experimental trace at 100 fs, and between *n*=5 and experimental trace at 600 fs.

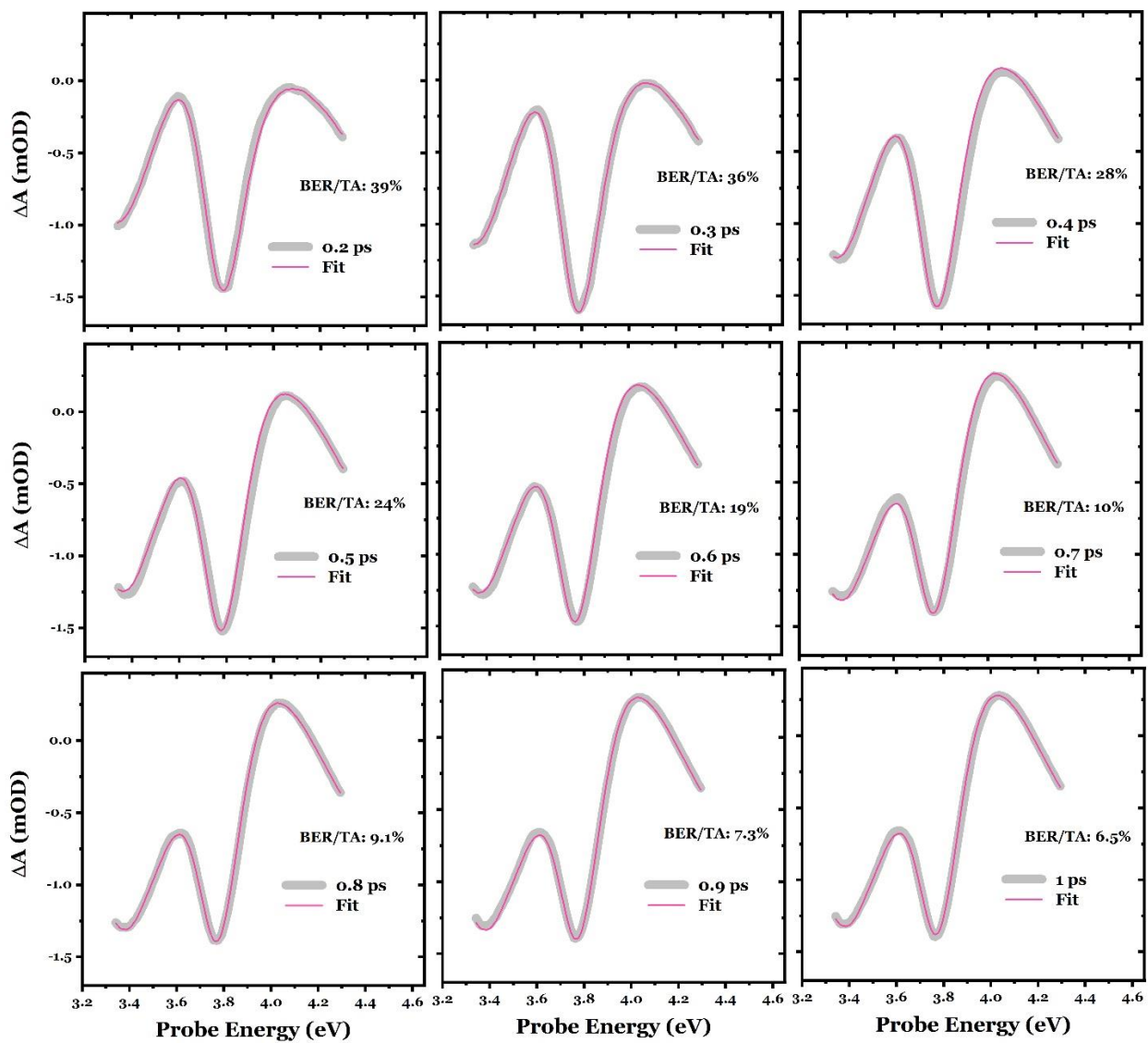

**Fig. S7.** Experimental UV-probed spectral traces and their fittings using different combinations of DAS spectra (a×DAS1 + b×DAS2). The fitting starting from 0.1 ps and with 0.1 ps intervals. The grey solid lines represent the experimental traces, while the magenta depict the fitted spectra. All figures share the same coordinate system. The relative contribution of BER to the overall TA signal (BER/TA) were demonstrated in each figure.

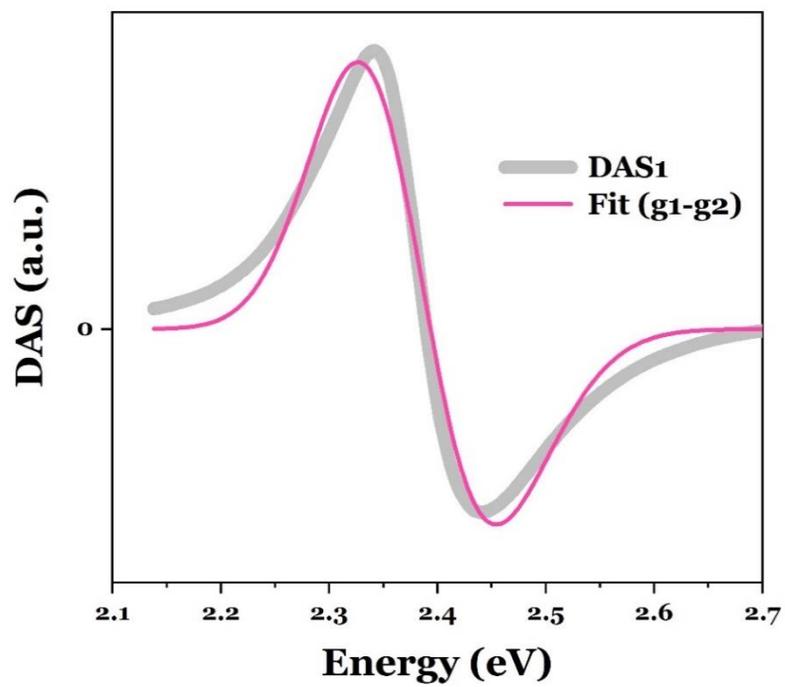

**Fig. S8.** Fitting of the DAS1 resulting from the visible probe signal, using two Gaussian functions (Gaussian 1 minus Gaussian 2). A band shift of 90 ± 30 meV was obtained.

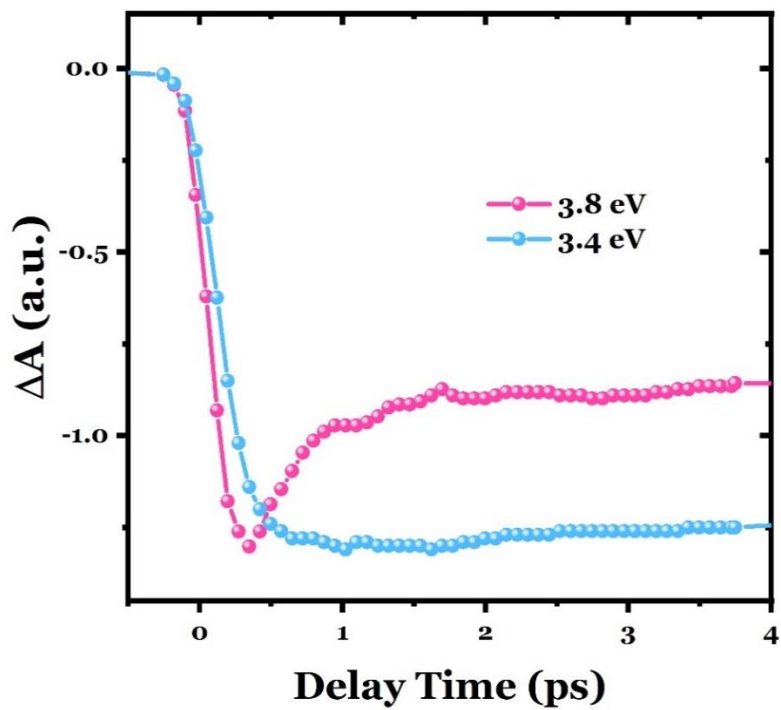

**Fig. S9.** Time traces of the rising TA signals probed at 3.4 and 3.8 eV, respectively. The traces are zoomed in to the first 4 ps and normalized at their maximum amplitudes.

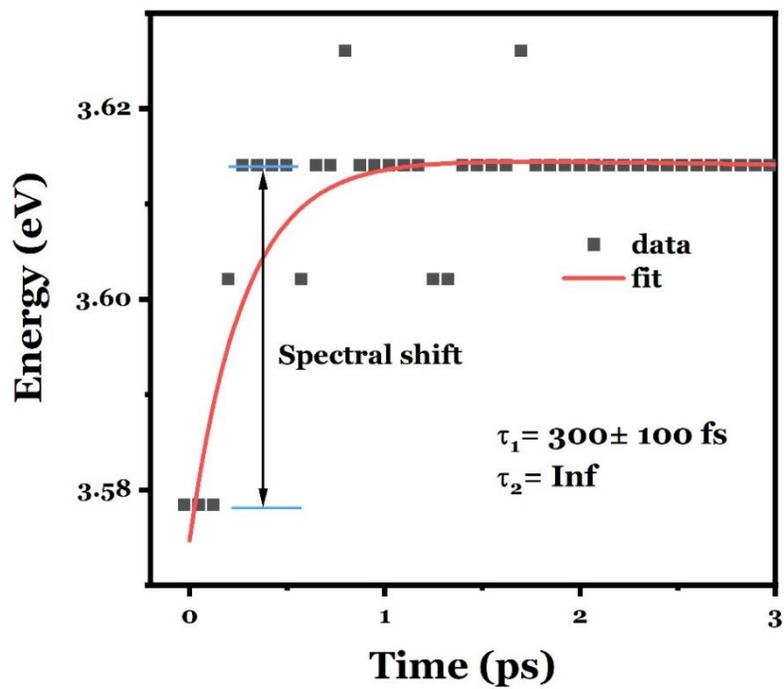

**Fig. S10.** Time evolution of the maximum peak at ~3.6 eV fitted with two exponential functions. The resulting first time constant is 300 ± 100 fs.